\documentclass[twoside,10pt]{article}
\usepackage{amsfonts,amsthm}

\catcode`\@=11 \long\def\@makefntext#1{ \protect\noindent \hbox to
3.2pt {\hskip-.9pt
$^{{\eightrm\@thefnmark}}$\hfil}#1\hfill}       

\def\@makefnmark{\hbox to 0pt{$^{\@thefnmark}$\hss}}    

\def\ps@myheadings{\let\@mkboth\@gobbletwo
\def\@oddhead{\hfill\hbox{}\rightmark}
\def\@oddfoot{}\def\@evenhead{\leftmark\hbox{}\hfill}\def\@evenfoot{}
\def\sectionmark##1{}\def\subsectionmark##1{}}
\oddsidemargin=\evensidemargin
\newcommand{\e}{\varepsilon}
\newcommand{\la}{\lambda}
\newcommand{\om}{\omega}
\newcommand{\D}{\Delta}
\newcommand{\dl}{\delta}
\newcommand{\ba}{\begin{array}{l}}
\newcommand{\ea}{\end{array}}
\font\tenrm=cmr10 \font\tenbf=cmbx10 \font\ninerm=cmr9
\font\eightrm=cmr8 \font\eightit=cmti8
\def\abstracts#1#2#3{{
    \centering{\begin{minipage}{4.5in}\footnotesize\baselineskip=10pt
    \parindent=0pt #1\par
    \parindent=15pt #2\par
    \parindent=15pt #3
    \end{minipage}}\par}}
\newcounter{sectionc}\newcounter{subsectionc}\newcounter{subsubsectionc}
\renewcommand{\section}[1] {\vspace{12pt}\addtocounter{sectionc}{1}
\setcounter{subsectionc}{0}\setcounter{subsubsectionc}{0}\noindent
    {\tenbf\thesectionc. #1}\par\vspace{5pt}}
\renewcommand{\subsection}[1] {\vspace{12pt}\addtocounter{subsectionc}{1}
    \setcounter{subsubsectionc}{0}\noindent
  {\bf\thesectionc.\thesubsectionc. {\kern1pt \bfit #1}}\par\vspace{5pt}}
\renewcommand{\subsubsection}[1] {\vspace{12pt}\addtocounter{subsubsectionc}{1}
    \noindent{\tenrm\thesectionc.\thesubsectionc.\thesubsubsectionc.
    {\kern1pt \tenit #1}}\par\vspace{5pt}}
\newcommand{\nonumsection}[1] {\vspace{12pt}\noindent{\tenbf #1}
    \par\vspace{5pt}}

\renewenvironment{thebibliography}[1]
    {\frenchspacing
     \ninerm
     \baselineskip=11pt
     \begin{list}{\arabic{enumi}.}
        {\usecounter{enumi}\setlength{\parsep}{0pt}
     \setlength{\leftmargin 12.7pt}{\rightmargin 0pt} 
         \setlength{\itemsep}{0pt} \settowidth
    {\labelwidth}{#1.}\sloppy}}{\end{list}}
\def\eightcirc{
\begin{picture}(0,0)
\put(4.4,1.8){\circle{6.5}}
\end{picture}}
\def\eightcopyright{\eightcirc\kern2.7pt\hbox{\eightrm c}}

\newcommand{\copyrightheading}[1]
    {\vspace*{-2.5cm}\baselineskip=10pt{\flushleft
    {\footnotesize Infinite Dimensional Analysis, Quantum Probability
          and Related Topics #1}\\
    {\footnotesize\copyright\kern2pt World Scientific Publishing
     Company}\\
     }}
\def\keywords#1{{
    \centering{\begin{minipage}{4.5in}\footnotesize\baselineskip=10pt
    {\footnotesize\it Keywords}\/: #1
     \end{minipage}}\par}}
\newtheorem{theorem}{Theorem}[sectionc] 
\newtheorem{lemma}{Lemma}[sectionc]
\newtheorem{remark}{Remark}[sectionc]
\newtheorem{proposition}{Proposition}[sectionc]
\def\qed{\hbox{${\vcenter{\vbox{         
   \hrule height 0.4pt\hbox{\vrule width 0.4pt height 6pt
   \kern5pt\vrule width 0.4pt}\hrule height 0.4pt}}}$}}

\begin{document}
\setlength{\textheight}{7.7truein}    

\markboth{\protect{\footnotesize\it Quantum Multipole Noise
}}{\protect{\footnotesize\it Quantum Multipole Noise}}

\normalsize\baselineskip=13pt\thispagestyle{empty}
\setcounter{page}{1}

\copyrightheading{}

\vspace*{1in}

\centerline{\tenbf QUANTUM MULTIPOLE NOISE AND} \baselineskip=13pt
\centerline{\tenbf GENERALIZED QUANTUM STOCHASTIC EQUATIONS}

\vspace*{0.37truein}

\centerline{\eightrm A. N. PECHEN\footnote{E-mail: pechen@mi.ras.ru}\,\,\,  and I. V.
VOLOVICH\footnote{E-mail: volovich@mi.ras.ru}}

 \baselineskip=12pt\centerline{\eightit Steklov Mathematical Institute, Russian Academy of Sciences,}
 \baselineskip=10pt\centerline{\eightit Gubkin St. 8, GSP-1, 117966, Moscow, Russia}

\vspace*{1truein}

\abstracts{\eightrm A notion of quantum multipole (in particular,
dipole) noise is considered. Quantum dipole noise is an analogue
of quantum white noise but it acts in a Fock space with indefinite
metric. Quantum {\eightit white} noise describes the leading term
in the stochastic limit approximation to quantum dynamics while
quantum {\eightit multipole} noise describes corrections to the
leading term. We obtain and study generalized quantum stochastic
equations describing corrections to the stochastic limit which
include quantum dipole noise.}{}{}

\vspace*{10pt} \keywords{multipole noise, white noise, dipole
noise, quantum stochastic equations}

\vspace*{4pt} \baselineskip=13pt \normalsize\tenrm
\section{Introduction}
\noindent \looseness1 There are numerous works devoted to the
studying of the long time behavior in quantum theory. Recently an
exact formula for the time dependence of certain matrix elements
(ABC-formula) has been obtained in~\cite{AV}.

The leading term in the weak coupling -- large time regime in this
formula corresponds to the stochastic limit (see~\cite{alv}). In
the operator formalism the stochastic limit is described by the
quantum white noise and corresponding quantum stochastic
differential equations~\cite{ALV}-\cite{Huang}.

To describe corrections to the stochastic limit one needs a new
operator structure. For free fields an appropriate new notion was
introduced in~\cite{VolBN}. We call it the quantum dipole noise.
In this paper we first study mathematical properties of the
quantum dipole noise and then consider interacting case and show
that the quantum dipole noise describes first order corrections to
the stochastic limit. We introduce quantum multipole noise which
describes corrections of any order to the stochastic limit.
Moreover we introduce a new type of quantum stochastic
differential equations which involves quantum dipole noise. These
quantum stochastic differential equations we call generalized
quantum stochastic differential equations.

Let us remind that the quantum white noise is an operator valued
distribution with commutation relations proportional to Dirac's $
\dl $--function. The simplest example of the quantum white noise
is given by the following commutation relations
\begin{equation}
 [b(t),b^+(\tau)]=\dl(\tau-t).
\end{equation}
The white noise operators are the subject of investigation in
numerous works \mbox{\cite{alv}-\cite{Huang}.}

In the present article we study higher order corrections to the
stochastic limit. We show that a new mathematical object: dipole
noise operators - naturally arises as the second order terms in
the expansion of collective operators $a^{\pm}(k,t/\la^2)$ in the
series with respect to the $ \la $. Here $ a^{\pm}(k,t):=e^{\pm
i\om(k)t}a^{\pm}(k) $, where $a^-(k)$ and $a^+(k)$ are bosonic
annihilation and creation operators.

An operator valued distribution with commutation relations
proportional to derivative of Dirac's $ \dl $--function has being
introduced in~\cite{VolBN}. We call it the quantum dipole noise.
The simplest example of quantum dipole noise is defined by the
commutation relations
\begin{equation}\label{sbnalg}
 [c(t),c^+(\tau)]=i\dl'(\tau-t).
\end{equation}

We introduce and study also further generalizations of the white
and dipole noise operators which we call multipole noise
operators. They satisfy the commutation relations
\begin{equation}
 [c_m(t),c^+_n(\tau)]=\dl_{n,m}i^n\dl^{(n)}(\tau-t),
\end{equation}
where $m,n=0,1,\dots$, $c_0(t)\equiv b(t)$ and $c_1(t)\equiv
c(t)$.

White noise operators arise as the first order terms in the
expansion of the free fields in the series with respect to $\la$
and given by the following limit (in the sense of convergence of
correlators, see~\cite{alv})
\begin{equation}
 \lim\limits_{\la\to 0}\frac{1}{\la}a^{\pm}(k,t/\la^2)=b^{\pm}(k,t).
\end{equation}
Here $ b^{\pm}(k,t) $ are the white noise operators. They satisfy
the following commutation relations
\begin{equation}\label{wncomm}
 [b^-(k,t),b^+(p,\tau)]=2\pi\dl(\om(k))\dl(\tau-t)\dl(p-k).
\end{equation}

Next terms of the expansion of the collective operators in the
series with respect to the $ \la $ have being found
in~\cite{VolBN}. They are the dipole noise operators $
c^{\pm}(k,t) $ satisfying the following commutation relations
\begin{equation}\label{bncomm}
  [c^-(k,t),c^+(p,\tau)]=2\pi i\dl'(\om(k))\dl'(\tau-t)\dl(p-k).
\end{equation}
We will also use notations $a(k)\equiv a^-(k)$, $b(k,t)\equiv
b^-(k,t)$ and $c(k,t)\equiv c^-(k,t)$.

In section~2 we will construct a representation of dipole
noise~(\ref{sbnalg}) in a pseudo-Hilbert space with indefinite
metric. We will show that the quantum dipole noise is a well
defined operator valued distribution in a Fock space with
indefinite metric. Notice that spaces with indefinite metric are
widely used in quantum field theory for quantization of
electromagnetic and other gauge fields (Gupta-Bleuler
formalism,~\cite{bogol2}, see also~\cite{AI,boloto}).

Using multipole (white and dipole) noise operators one can make
the formal expansion
\begin{equation}\label{expansion}
a^{\pm}(k,t/\la^2)=\la b^{\pm}(k,t)+\la^2 c^{\pm}(k,t)+o(\la^2).
\end{equation}
The base for this expansion is the following expansion (in the
sense of distributions)
$$
\frac{1}{\la^2}e^{itx/\la^2}=2\pi\dl(x)\dl(t)+\la^22\pi
i\dl'(x)\dl'(t)+o(\la^2),
$$
where $x$ and $t$ are real variables.

Using expansion~(\ref{expansion}) of the collective (in this case
free) operators we obtain the equation for the higher order
corrections to the stochastic limit of the evolution operator $
U_\la(t)$ of system interacting with Bose field. Then we bring
this equation to the normally ordered form which is convenient for
calculation of corrections for matrix elements of the evolution
operator.

One has the expansion for the rescaled evolution operator into the
series at the coupling constant $\la$
$$
 U_\la(t/\la^2)=U_0(t)+\la U_1(t)+\dots .
$$

The normal form of the equation for the evolution operator
$U_0(t)$ in the stochastic limit is equivalent to the quantum
stochastic differential equation in the sense of~\cite{HP} and
looks like
\begin{equation}\label{dU0(t)}
 dU_0(t)=(DdB^+_t-D^+dB_t-\gamma D^+Ddt)U_0(t).
\end{equation}
Here we use standard notations $ dB_t $ and $ dB^+_t $ for
stochastic differentials of quantum Brownian motion; $ D $ and $
D^+ $ are bounded operators in system Hilbert space and $ \gamma $
is a complex number.

We will obtain quantum differential equations of a new form which
include quantum multipole noise. They will describe the higher
order corrections to the stochastic limit. For example for the
first order correction $U_1(t)$ to the evolution operator one has
the generalized quantum stochastic differential equation
\begin{equation}\label{dU1(t)}
 dU_1(t)=(DdC^+_t-D^+dC_t)U_0(t)+(DdB^+_t-D^+dB_t-\gamma
 D^+Ddt)U_1(t).
\end{equation}
We call it generalized quantum stochastic differential equation
because it contains stochastic differentials of dipole noise which
we denoted as $ dC_t $ and $ dC^+_t $. These operators are defined
in a Fock space with indefinite metric.

In Sect.~2 quantum dipole noise operators in the Fock space with
indefinite metric are defined. An expansion of the free fields
into the series containing multipole noise operators is obtained
in Sect.~3. Dipole noise on the simplex is considered in Sect.~4
wcorrections to the Green function are described in Sect.~5.
Finally generalized quantum stochastic differential equations
describing corrections to the stochastic limit are presented in
Sect.~6-9.

\section{Quantum Dipole Noise in Fock Space with Indefinite Metric}\label{sectbn}
\noindent The commutation relations~(\ref{sbnalg}) for the dipole
noise mean that one has a family of operators $ c_g $
parameterized by complex valued functions of real argument $ g(t)
$ from some space of functions (for example from Schwartz space of
functions $S({\mathbb R})$) with commutation relations
\begin{equation}\label{bnalg}
 [c_f,c^+_g]=<f,g>,
\end{equation}
where
\begin{equation}\label{sform}
 <f,g>=i\int\overline{f'(t)}g(t)dt.
\end{equation}
Notice that the inner product $ <\cdot,\cdot> $ is non-positively defined, i.e. it is an indefinite
inner product. In fact, for $ f(t)=u(t)+iv(t) $ with $ u(t) $ and $ v(t) $ are real valued
functions one has
$$
 <f,f>=-2\int u'(t)v(t)dt.
$$
This quantity can be positive or negative. Therefore the natural
space of representation of algebra~(\ref{bnalg}) is a vector space
with indefinite inner product.

Let us describe a general construction of a Fock space with indefinite metric
(see~\cite{AI,boloto}). Let $ {\cal H}_1 $ be a Hilbert space over field ${\mathbb C}$ and
$(\cdot,\cdot)_{{\cal H}_1}$ is the inner product in $ {\cal H}_1 $. This inner product induces in
${\cal H}_1$ a structure of complex Fre\'chet space with Hilbert topology. An indefinite inner
product on $ {\cal H}_1 $ is a hermitian sesquilinear form, i.e. a map $ {\cal H}_1\times{\cal
H}_1\to {\mathbb C} $ with the properties
$$
 <f,\alpha g+\beta h>=\alpha<f,g>+\beta<f,h>
$$
$$
 <f,g>=\overline{<g,f>}
$$
for all $ f,g,h\in {\cal H}_1 $ and $\alpha,\beta\in {\mathbb C}
$. The form $<\cdot,\cdot>$ is non necessarily positively defined.
It is called an indefinite metric. We suppose  that if $F(\cdot)$
is a continuous linear functional over the topological space $
{\cal H}_1 $ then it can be uniquely represented in the form
\begin{equation}\label{F(g)}
 F(g) = <f, g>,\qquad g\in {\cal H}_1
\end{equation}
for some $f\in {\cal H}_1$. Moreover, we suppose that for any $f\in {\cal H}_1$ Eq.~(\ref{F(g)})
determines a linear continuous functional $F(g)$ over ${\cal H}_1$. If~(\ref{F(g)}) holds then we
say that the Hilbert topology is consistent with the indefinite inner product. In this case the
complex Fre\'chet space with Hilbert topology and with the indefinite metric is being called a
pseudo-Hilbert space. The condition of~(\ref{F(g)}) is equivalent to existence of a bounded
hermitian with respect to the inner product $(\cdot,\cdot)_{{\cal H}_1}$ linear operator $\eta:
{\cal H}_1\to {\cal H}_1$ such that the inverse operator $\eta^{-1}$ exists, bounded and moreover
$\eta$ satisfies the following property:
$$
 <f,g>=(f,\eta g)_{{\cal H}_1},\qquad\forall f, g\in {\cal H}_1.
$$

Notice that one can vary the inner product in the definition of
Hilbert topology. In fact, two different inner products in ${\cal
H}_1$, say $(\cdot,\cdot)_{{\cal H}_1}$ and $(\cdot,\cdot)'_{{\cal
H}_1}$, induce the same topology if $\exists\,\, C,\,C' > 0$:
$$
 C(f,f)_{{\cal H}_1}\leq (f,f)'_{{\cal H}_1}\leq C'(f,f)_{{\cal
 H}_1}.
$$
Moreover, one can choose the inner product in such a way that the
operator $\eta$ satisfies an additional condition
\begin{equation}\label{eta2}
 \eta^2=1.
\end{equation}
We will denote in this paper as $(\cdot,\cdot)_{{\cal H}_1}$ the
inner product for which operator $\eta$ satisfies~(\ref{eta2}).

Let $ {\cal H}_1^{\otimes^n_s}:={\rm Sym\ }{\cal
H}_1\otimes\dots\otimes{\cal H}_1 $ be a symmetric tensor product
of $n$ copies of ${\cal H}_1$ for $n\ge 1$ and ${\mathbb C}$ for
$n=0$. Operator of symmetrization acts on $ \tilde
f_n=f^{(1)}\otimes\dots\otimes f^{(n)} $ as
\begin{equation}\label{fn}
 {\rm Sym\
 }\tilde f_n=\frac{1}{n!}\sum\limits_\pi f^{(\pi(1))}\otimes\dots\otimes f^{(\pi(n))}
 =:f_n
\end{equation}
with summation over all permutations of the set $ \{1,\dots,n\}$.
The indefinite inner product of two vectors $f_n,\, g_n\in{\cal
H}_1^{\otimes^n_s}$ for $n>0$ is given by
$$
 <f_n, g_n>:=(f_n, \eta^{\otimes^n}g_n)_{{\cal
H}_1^{\otimes^n_s}}
$$
where $\eta^{\otimes^n}:=\eta\otimes\dots\otimes\eta$ is the
tensor product of $n$ copies of $\eta$. For $n=0$ we define
$$
 <f_0,g_0>:=\bar f_0g_0
$$

Then we can define the Boson Fock space with indefinite metric
$$
 {\cal F}({\cal H}_1)=\bigoplus\limits_{n=0}^{\infty}{\cal
 H}_1^{\otimes^n_s}.
$$
We mean that any vector $\phi\in {\cal F}({\cal H}_1) $ is a
sequence of the form
$$
 \phi=(f_0,f_1,\dots,f_n,\dots)
$$
with the convergent series
$$
 ||\phi||^2=\sum\limits_{n=0}^\infty (f_n,f_n)_{{\cal
 H}^{\otimes^n_s}_1} < \infty.
$$
Here $ f_0\in {\mathbb C} $ is a complex number, $ f_n $ given
by~(\ref{fn}) and is being called the $ n $--particle component of
the vector $\phi$. The indefinite inner product of two vectors
$\phi=(f_0,f_1,\dots)$ and $\psi=(g_0,g_1,\dots)$ is given by
$$
 <\phi,\psi>=\sum\limits_{n=0}^\infty<f_n,g_n>.
$$
The series is convergent because $<f_n,g_n>=(f_n,\eta^{\otimes^n}
g_n)_{{\cal H}_1}$ and $\eta^2=1$.

Creation and annihilation operators  $ c^+_f $ and $ c_f $ act on $n$--particle component of the
vector $ \phi $ by usual formulae
$$
 (c^+_ff_n)_{n+1}=\sqrt{n+1}{\rm\ Sym\ }f\otimes f_n
$$
$$
 (c_ff_n)_{n-1}=\frac{1}{\sqrt{n}}\sum\limits_{i=1}^n<f,f^{(i)}>{\rm
 \ Sym\ }f^{(1)}\otimes\dots\hat f^{(i)}\dots\otimes f^{(n)}.
$$
Here $ \hat f^{(i)} $ means that $ f^{(i)} $ is missed. One has
the canonical commutation relations (CCR)
$$
[c_f,c^+_g]=<f,g>
$$
and also on finite vectors one has the relation
$$
 <c_f\phi,\psi>=<\phi,c^+_f\psi>
$$
which means that the annihilator $c_f$ is the adjoint of the
creator $c^+_f$ with respect to the inner product $<\cdot,\cdot>$.

In order to construct a representation of algebra~(\ref{bnalg})
let us choose $ {\cal H}_1 $ to be the completion of the Schwartz
space $S({\mathbb R})$ with respect to inner product
\begin{equation}\label{HilbTop}
 (f,g)_{{\cal H}_1}: = \frac{1}{2\pi}\int |\tau|\overline{\tilde f(\tau)}\tilde
 g(\tau)d\tau.
\end{equation}
Here
$$
 \tilde f(\tau)=\int e^{ix\tau}f(x)dx
$$
is the Fourier transform of $ f(x) $. We introduce an indefinite
inner product on $ {\cal H}_1 $ by the formula
\begin{equation}\label{IndM}
 <f,g>\ = i\int\overline{f'(t)}g(t)dt
\end{equation}
which can be rewritten in the form
$$
 <f,g>\ = \frac{1}{2\pi}\int\tau\overline{\tilde f(\tau)}\tilde
 g(\tau)d\tau.
$$
In this case the operator $\eta$ acts on the Fourier transform of
the function $f(x)$ as
$$
 (\eta\tilde f)(\tau)={\rm sign}(\tau)\tilde f(\tau)
$$
where ${\rm sign}(\tau)=1$ if $\tau\ge 0$ and $-1$ in the opposite
case. Clearly this operator satisfies~(\ref{eta2}) and all
conditions which are necessary to guarantee~(\ref{F(g)}).
Therefore the Hilbert topology which given by~(\ref{HilbTop}) is
consistent with the indefinite metric~(\ref{IndM}).

Any vector $ \phi\in {\cal F}({\cal H}_1) $ is a sequence of
(symmetric) functions,
$$
 \phi=(f_0,f_1(t_1),\dots,f_n(t_1,\dots,t_n),\dots).
$$
Here $ f_0\in {\mathbb C} $ is a complex number as above.

According to general construction described above creation and
annihilation dipole noise operators $ c^+_f $ and $ c_f $ act on
$n$-particle component of the vector $ \phi $ by formulae
$$
 (c^+_ff_n)_{n+1}(t_1,\dots,t_{n+1})=\frac{1}{\sqrt{n+1}}
 \sum\limits_{i=1}^{n+1}f(t_i)f_n(t_1,\dots,\hat
 t_i,\dots,t_{n+1}).
$$
$$
 (c_ff_n)_{n-1}(t_1,\dots,t_{n-1})=i\sqrt{n}\int\overline{f'(t)}
 f_n(t,t_1,\dots,t_{n-1})dt
$$
Here $ \hat t_i $ means that argument $ t_1 $ is missed.

\begin{remark}\normalsize\tenrm
Motivated by the investigation of the space with the sesquilinear form~(\ref{sform}) let us note
that it is interesting to consider the following approach to solution of the Dirichlet problem. Let
us consider the Laplace equation
\begin{equation}\label{laplace}
-\D u(x) = f(x),\quad x\in\Omega\subset\mathbb R^d
\end{equation}
where $\Omega$ is an arbitrary domain in $\mathbb R^d$. Denote $\cal H$ the Hilbert space obtained
after the completion and taking the factor space of the space of test functions with compact
support ${\cal D}(\Omega)$ with the scalar product
$$
 (\varphi,\psi)_{\cal H} = \int\limits_\Omega\nabla\bar\varphi\nabla\psi dx.
$$
We write Eq.~(\ref{laplace}) in the form
\begin{equation}\label{laplace2}
(u,\varphi)_{\cal H} = (f,\varphi),\qquad\varphi\in{\cal D}(\Omega)
\end{equation}
where
$$
 (f,\varphi) = \int\limits_\Omega\overline{f(x)}\varphi(x)dx.
$$

\begin{proposition}
Let $f\in L_{loc}(\Omega)$ and there exists a constant $C_\Omega > 0$ such that
$$
 |(f,\varphi)|^2\leq C_\Omega(\varphi,\varphi)_{\cal H}
 \qquad\forall\varphi\in{\cal D}(\Omega).
$$
Then there exists a unique solution $u\in\cal H$ of Eq.(\ref{laplace2}).
\end{proposition}
\end{remark}


\section{Multipole Noise}\label{sectionBN}\noindent
In this section we will show that the quantum multipole noise
naturally arises from studying of the higher order corrections to
the stochastic limit.

Let $L^2({\mathbb R}^d)$ be the Hilbert space (with inner product
$(\cdot,\cdot)$) of square integrable on ${\mathbb R}^d$
functions. Denote then
$$
{\cal F}=\bigoplus_{n=0}^\infty L^2({\mathbb R}^d)^{\otimes^n_s}
$$
the Boson Fock space and $a(f), a^+(g)$ are annihilation and
creation operators in $\cal F$ with commutation relations
\begin{equation}\label{aa+}
 [a(f),a^+(g)]=(f,g).
\end{equation}
Here $f,\ g\in L^2({\mathbb R}^d)$ and we will use the notation
$$
 a(f)=\int \bar f(k)a(k)dk,\qquad k\in\mathbb R^d.
$$

For a given complex valued test function $ g(k) $ on $ {\mathbb
R}^d $ and for smooth real valued function $ \om(k) $ define the
(smeared) time dependent creation and annihilation operators as
\begin{equation}\label{collop}
 A^+(t)=\int g(k)e^{i\om(k)t}a^+(k)dk,\qquad
 A(t)=\int \bar g(k)e^{-i\om(k)t}a(k)dk.
\end{equation}
The test function $ g(k) $ in physical applications plays a role
of formfactor describing structure of interaction. The function $
\om(k) $ has meaning of dispersion law so that these operators are
free evolutions of smeared creation and annihilation operators. We
assume that the surface $\om(k)=0$ is nondegenerate, i.e. ${\rm
grad}\, \om\ne 0$ on the surface $\om(k)=0$.

Let us consider the commutator of rescalled creation and
annihilation operators
$$
 A_\la(t):=\frac{1}{\la}A(t/\la^2),\qquad
 A^+_\la(t):=\frac{1}{\la}A^+(t/\la^2)
$$
(we call them collective operators), where $\la$ is a real
parameter. One has
$$
\left[A_\la(t), A^+_\la(\tau)\right]= \frac{1}{\la^2}\int
dk|g(k)|^2e^{i\frac{\om(k)(\tau-t)}{\la^2}}.
$$
In order to investigate the behavior of this commutator for small
$ \la $ let us consider the sequence of functions on $ {\mathbb
R^2} $ indexed by real number $ \la\ne 0 $
$$
f_{\la}(x,t)=\frac{1}{\la^2}e^{ixt/\la^2} .
$$
For any $\la\ne 0$ $f_\la(x,t)$ is locally integrable. Therefore it defines a distribution from
$S'({\mathbb R}^2)$. In the sense of distributions one has the relation~(\cite{alv})
$$
\lim\limits_{\la\to 0}f_{\la}(x,t)=2\pi\dl(x)\dl(t).
$$
We will prove the following generalization of this relation.

\begin{theorem}\label{limit1}
One has the asymptotic expansion (in the sense of distributions)
\begin{equation}\label{asympt}
\frac{1}{\la^2}e^{ixt/\la^2}=2\pi\sum\limits_{n=0}^\infty
\frac{(i\la^2)^n}{n!}\dl^{(n)}(x)\dl^{(n)}(t).
\end{equation}
The asymptotic series~(\ref{asympt}) means that for any test
functions $f$ and $\phi$ from the Schwartz space $S({\mathbb R})$
and for any $N\in\mathbb N$
$$
\int\frac{1}{\la^2}e^{ixt/\la^2}f(x)\phi(t)dxdt-2\pi\sum\limits_{n=0}^N
\frac{(i\la^2)^n}{n!}f^{(n)}(0)\phi^{(n)}(0)=o(\la^{2N}).
$$
\end{theorem}

\noindent{\bf Proof.} Let $ f(x) $ and $ \phi(t) $ be test
functions from the Schwartz space $S({\mathbb R})$. Then one has
$$
\int\frac{1}{\la^2}e^{ixt/\la^2}f(x)\phi(t)dxdt = \int\tilde
f(\tau)\phi(\la^2\tau)d\tau.
$$
Here
$$
 \tilde f(\tau)=\int e^{ix\tau}f(x)dx
$$
is the Fourier transform of $ f(x) $. It is well-known
(see~\cite{vladimirov}) that $ \tilde f(\tau)\in S({\mathbb R}) $
and the inverse Fourier transform is given by the formula
$$
 f(x)=\frac{1}{2\pi}\int e^{-ix\tau}\tilde f(\tau)d\tau.
$$

Since $\phi(t)\in S(\mathbb R)$ one can write the Taylor expansion
for $\phi(\la^2\tau)$ in the Lagrange form
$$
\phi(\la^2\tau)=\phi(0)+\sum\limits_{n=1}^N\frac{(\la^2\tau)^n}{n!}\phi^{(n)}(0)+
\frac{(\la^2\tau)^{N+1}}{(N+1)!}\phi^{(N+1)}(\xi_{\la}(\tau))
$$
where $\xi_{\la}(\tau)\in (0,\la^2\tau)$. Using this expansion one
gets
$$
\int\tilde f(\tau)\phi(\la^2\tau)d\tau-\sum\limits_{n=0}^N
\frac{\la^{2n}}{n!}\phi^{(n)}(0)\int\tau^n\tilde f(\tau)d\tau
$$
\begin{equation}\label{th1}
=\frac{\la^{2(N+1)}}{(N+1)!}\int\tau^{N+1}\tilde
f(\tau)\phi^{(N+1)}(\xi_\la(\tau))d\tau.
\end{equation}

Now notice that
$$
\int\tau^n\tilde f(\tau)d\tau=2\pi i^nf^{(n)}(0).
$$
Then LHS of~(\ref{th1}) becomes equal to
$$
\int\frac{1}{\la^2}e^{ixt/\la^2}f(x)\phi(t)dxdt-2\pi\sum\limits_{n=0}^N
\frac{(i\la^2)^n}{n!}f^{(n)}(0)\phi^{(n)}(0).
$$
It is clear that RHS of~(\ref{th1}) is
$o(\la^{2N})$.\hspace{6cm}\qed

\begin{remark}\label{remark1}\normalsize\tenrm
Let us denote non smeared creation and annihilation collective operators as
$$
 a_\la^\pm(k,t):=\frac{1}{\la}a^\pm(k,t/\la^2)
$$
so that
$$
 A^+_\la(t)=\int g(k)a^+_\la(k,t)dk.
$$
The expansion~(\ref{asympt}) for $ f_\la(x,t) $, which can be
symbolically rewritten in the form
$$
 \frac{1}{\la^2}e^{itx/\la^2}=2\pi\exp\left(i\la^2
 \frac{\partial^2}{\partial x\partial t}\right)\dl(x)\dl(t),
$$
induces the corresponding expansion for the non smeared collective operators
\begin{equation}\label{exp3}
 a_\la(k,t)=\sum\limits_{n=0}^N\la^nc_n(k,t)+o(\la^N)
\end{equation}
such that $c_n(k,t),\ n=0,1,\dots$ satisfy the commutation
relations
$$
 [c_n(k,t),c^+_m(p,\tau)]=\dl_{n,m}\frac{2\pi
 i^n}{n!}\dl^{(n)}(\om(k))\dl^{(n)}(\tau-t)\dl(k-p).
$$
\end{remark}

\begin{remark}\normalsize\tenrm
It would be interesting to study in that sense one can speak about
a spectral decomposition of initial Fock space $\cal F$ into a
family of spaces corresponding to the multipole noise creation and
annihilation operators.
\end{remark}

We will use also the notations
$$
 c_0(k,t)\equiv b(k,t),\qquad c_1(k,t)\equiv c(k,t),\qquad c_2(k,t)\equiv d(k,t).
$$
Here $ b(k,t), b^+(k,t) $ are the white noise operators:
$$
 [b(k,t),b^+(p,\tau)]=2\pi\dl(\om(k))\dl(\tau-t)\dl(k-p)
$$
and $ c(k,t), c^+(k,t) $ are the dipole noise operators:
$$
 [c(k,t),c^+(p,\tau)]=2\pi i\dl'(\om(k))\dl'(\tau-t)\dl(k-p).
$$
The second order terms in the expansion~(\ref{exp3}) (we denote
their as $ d(k,t) $ and $ d^+(k,t) $) satisfy the commutation
relations
$$
 [d(k,t),d^+(p,\tau)]=-\pi\dl''(\om(k))\dl''(\tau-t)\dl(k-p).
$$

For the smeared time dependent operators one has an expansion
\begin{equation}\label{exp4}
 A(t/\la^2)=\sum\limits_{n=0}^N\la^{n+1}c_n(t)+o(\la^{N+1}).
\end{equation}
The operators $c_n(t)$ satisfy the commutation relations
$$
 [c_n(t),c^+_m(\tau)]=\dl_{n,m}\tilde\gamma_n\dl^{(n)}(\tau-t)
$$
with complex numbers
$$
 \tilde\gamma_n=\frac{(-1)^n}{n!}
 \int\limits_{-\infty}^{\infty}\sigma^nd\sigma\int
 e^{i\sigma\om(k)}|g(k)|^2dk.
$$

The expansion~(\ref{exp4}) holds for any $N\in\mathbb N$ in the
case then the set of stationary points of function $\om(k)$ has
zero intersection with ${\rm supp}\ g(k)$, i.e in the case
$|g(k)|^2\in S(\Omega)$ for any $\Omega\in \mathbb R^d$ such that
stationary points of $\om(k)$ do not belong to $\Omega$. The set
of stationary points of $\om(k)$ is defined as ${\cal
K}:=\{k_0\,\,|\,\, {\rm grad}\,\om(k_0)=0\}$. The
expansion~(\ref{exp4}) holds for $N=1$ and $|g(k)|^2\in S(\mathbb
R^d)$ for $d\ge 3$.


\section{Dipole Noise on the Simplex}\label{sectionSim}\noindent
Some modification of Theorem~\ref{limit1} is useful for getting
the differential equation for the first order correction. This
modification is about distributions on the standard simplex
(see~\cite{alv}).

Let us remind the construction for distributions on the standard
simplex. Define
$$
 C_0:=\{\phi:\mathbb R_+\to\mathbb C\ |\ \phi=0 {\rm\ a.e.}\}
$$
$$
 \tilde C_1:=\{\phi:\mathbb R_+\to\mathbb C\ |\ \phi
 {\rm\ is\ bounded\ and\ left-continuous\ at\ any\ } t>0\}
$$
$$
 \tilde C:={\rm\ linear\ span\ of\ } \{C_0\cup \tilde C_1\}.
$$
For any $ a>0 $ define $\dl_+(\cdot-a) $ as the unique linear
extension of the map
$$
 \dl_+(\cdot-a):\phi\in \tilde C_1\to\phi(a)
$$
$$
 \dl_+(\cdot-a):\phi\in C_0\to 0.
$$

One has the following result: in the sense of distributions on the
simplex there exists the limit (see~\cite{alv} for more
discussions)
\begin{equation}\label{dplus}
 \lim\limits_{\la\to 0}\frac{1}{\la^2}e^{ix(t-a)/\la^2}=
 \frac{1}{i(x-i0)}\dl_+(t-a).
\end{equation}

In order to introduce the derivative $ \dl'_+(\cdot-a) $ we will
define $ C_1 $ as
$$
 C_1:=\{\phi:\mathbb R_+\to\mathbb C\ |\
  \phi(t)=\sum\limits_{i=1}^n\phi_i(t)\Theta_{[0,a_i]}(t)\ {\rm for\
  some\ } n\in\mathbb N, 0<a_i\le\infty\ {\rm and\ }
$$
$$
  \phi_i(t)\ {\rm is\ bounded,\ with\ bounded\ and\ continuous\ first\ derivative\ }\phi_i'(t)
  \}.
$$
This definition means that for any $\phi\in C_1$ and $a>0$ left--derivative
$$
 \phi'_L(a):=\lim\limits_{\varepsilon\to -0}\frac{\phi(a+\varepsilon)-\phi(a)}
  {\varepsilon}
$$
exists. Moreover it is left--continuous and bounded.

With such defined $C_1$ we define $C$ as linear span of $\{C_0\cup C_1\}$. Clearly $C\subset \tilde
C$. We will denote the restriction of $\dl_+(\cdot-a)$ to $C$ again as $\dl_+(\cdot-a)$.

For any $ a>0 $ define $ \dl'_+(\cdot-a) $ as the unique linear
extension of the map
$$
 \dl_+'(\cdot-a):\phi\in C_1\to-\phi'_L(a)
$$
$$
 \dl_+'(\cdot-a):\phi\in C_0\to 0.
$$

With such defined $ \dl'_+(\cdot-a) $ the first order correction
to~(\ref{dplus}) (proportional to $\la^2$) is given by the
\begin{theorem}\label{limit2}
One has the following limit in the sense of distributions on the
standard simplex ($ a>0 $ is a positive number)
$$
 \lim\limits_{\la\to 0}\frac{1}{\la^2}\left(\frac{1}{\la^2}e^{ix(t-a)/\la^2}-
  \frac{1}{i(x-i0)}\dl_+(t-a)\right)=\dl'_+(t-a)\frac{d}{dx}\frac{1}{x-i0}.
$$
This means that for arbitrary test functions $ f(x)\in S(\mathbb
R) $ and $ \phi(t)\in C $ one has the limit
$$
 \lim\limits_{\la\to 0}\frac{1}{\la^2}\left[\frac{1}{\la^2}
 \int\limits_{-\infty}^{\infty}dx\int\limits_0^adt
 f(x)\phi(t)e^{ix(t-a)/\la^2}-
 (\dl_+(\cdot-a),\phi)\int\limits_{-\infty}^0dy\tilde f(y)\right]
$$
\begin{equation}\label{dpl}
 =-(\dl'_+(\cdot-a),\phi)\int\limits_{-\infty}^0dyy\tilde f(y).
\end{equation}
Here
$$
 \tilde f(y)=\int e^{ixy}f(x)dx.
$$
\end{theorem}

\noindent{\bf Proof.} The Eq.~(\ref{dpl}) clearly holds for
$\phi\in C_0$. Therefore let us consider the case $\phi\in C_1$.
It is enough to consider $\phi$ of the form $
\phi(x)=\phi_1(x)\Theta_{[0,a_1]}(x) $ with $ \phi_1 $ bounded,
with bounded and continuous first derivative. In this case
$(\dl_+(\cdot-a),\phi)= \phi(a)$ and
$(\dl_+'(\cdot-a),\phi)=-\phi'_L(a)$. Making change of the
variable $ y=(t-a)/\la^2 $ in the first integral one gets
$$
 L:=\lim\limits_{\la\to 0}\frac{1}{\la^2}
 \left[\int\limits_{-\infty}^{\infty}dx
 \int\limits_{-a/\la^2}^0dyf(x)\phi(\la^2y+a)e^{ixy}-
 \phi(a)\int\limits_{-\infty}^0dy\tilde f(y)\right]=
$$
$$
 \lim\limits_{\la\to 0}\frac{1}{\la^2}\left[
 \int\limits_{-a/\la^2}^0dy\tilde f(y)\phi(\la^2y+a)-
 \phi(a)\int\limits_{-\infty}^0dy\tilde f(y)\right].
$$

Let us consider the case $a>a_1$. In this case $\phi(a)=0$
and function $\phi(\la^2y+a)$ is equal
to zero outside of the interval $[-a/\la^2,(a_1-a)/\la^2]$. Therefore one has
$$
 L=\lim\limits_{\la\to 0}\frac{1}{\la^2}\int\limits_{-a/\la^2}^{(a_1-a)/\la^2}
 dy\tilde f(y)\phi_1(\la^2y+a)=0
$$
because of boundness of $\phi_1$ and Lemma~\ref{lemma}. In this case $\phi'_L(a)=0$ so RHS
of~(\ref{dpl}) also equal to zero and~(\ref{dpl}) holds.

Now let us consider the case $a\le a_1$. In this case $\phi(a)=\phi_1(a)$ and one has
$$
 L=\lim\limits_{\la\to 0}\frac{1}{\la^2}\left[
 \int\limits_{-a/\la^2}^0dy\tilde f(y)\phi_1(\la^2y+a)-
 \phi_1(a)\int\limits_{-\infty}^0dy\tilde f(y)\right]=
$$
$$
 \lim\limits_{\la\to 0}\frac{1}{\la^2}\left[
 \int\limits_{-\infty}^0dy\tilde f(y)\phi_1(\la^2y+a)-
 \int\limits_{-\infty}^{-a/\la^2}dy\tilde f(y)\phi_1(\la^2y+a)-
 \phi_1(a)\int\limits_{-\infty}^0dy\tilde f(y)\right]=
$$
$$
 \lim\limits_{\la\to 0}\frac{1}{\la^2}\left[
 \int\limits_{-\infty}^0dy\tilde f(y)\left(\phi_1(\la^2y+a)-\phi_1(a)\right)-
 \int\limits_{-\infty}^{-a/\la^2}dy\tilde
 f(y)\phi_1(\la^2y+a)\right].
$$
Using Lemma~\ref{lemma} we get
$$
 L=\lim\limits_{\la\to 0}\frac{1}{\la^2}\left[\int\limits_{-\infty}^0
 dy\tilde f(y)\left(\phi_1(\la^2y+a)-\phi_1(a)\right)\right].
$$
Using Lagrange's theorem one gets
$$
 \phi_1(\la^2y+a)-\phi_1(a)=\la^2y\phi'_1(\xi_\la(y))
$$
with $\xi_\la(y)\in (a+\la^2y, a)$. Therefore
$$
 L=\lim\limits_{\la\to 0}\int\limits_{-\infty}^0
 dyy\tilde f(y)\phi'_1(\xi_\la(y)).
$$
Now since $\phi'_1(\xi_\la(y))$ is bounded and $\tilde f(y)\in S(\mathbb R)$ one can apply
Lebesgue's dominated convergence theorem. Finally one has
$$
 L=\phi'_1(a)\int\limits_{-\infty}^0y\tilde f(y)dy=
 \phi'_L(a)\int\limits_{-\infty}^0y\tilde f(y)dy.\qed
$$

\begin{lemma}\label{lemma}
For arbitrary test function $f(x)\in S({\mathbb R})$ and two positive numbers $ b>a>0 $ one has
$$
  \int\limits_{a/\la^2}^{b/\la^2}dxf(x)=o(\la^2).
$$
\end{lemma}

\noindent{\bf Proof.} In fact for large enough $x$ and $\forall
n\in {\mathbb N}$ $ \exists C=Const $ such that $
|f(x)|\le\frac{C}{x^n}$. Let us choose $ n=3 $. Then one has
$$
 \lim\limits_{\la\to 0}\frac{1}{\la^2}\left|\,
 \int\limits_{a/\la^2}^{b/\la^2}dxf(x)\right|\le
 \lim\limits_{\la\to 0}\frac{1}{\la^2}\int\limits_{a/\la^2}^{b/\la^2}dx|f(x)|\le
 \lim\limits_{\la\to 0}\frac{C}{\la^2}\int\limits_{a/\la^2}^{b/\la^2}
 \frac{dx}{x^3}=0.\hspace{2.2cm}\qed
$$

\begin{remark}\label{remdlp}\normalsize\tenrm
We get the following relation
$$
 \frac{1}{\la^2}e^{ix(t-a)/\la^2}=\dl_+(t-a)\frac{1}{i(x-i0)}+
   i\la^2\dl'_+(t-a)\frac{d}{dx}\frac{1}{i(x-i0)}+o(\la^2).
$$
The corresponding expansion for the collective operators has the
form
$$
 A(t/\la^2)=\la b(t)+\la^2c(t)+o(\la^2)
$$
with causal commutation relations
$$
 [b(t),b^+(\tau)]=\gamma_0\dl_+(\tau-t)
$$
$$
 [c(t),c^+(\tau)]=\gamma_1\dl'_+(\tau-t).
$$
Here complex numbers $\gamma_n$ are given by
$$
 \gamma_n=\frac{(-1)^n}{n!}
 \int\limits_{-\infty}^0\sigma^nd\sigma\int
 e^{i\sigma\om(k)}|g(k)|^2dk.
$$
\end{remark}


\section{Corrections to the Green's Functions}\noindent
The master field in the standard formulation of the stochastic
limit is defined as the limit as $ \la\to 0 $ of the Wightman
correlation functions:
$$
  <0|A^1_{\la}(t_1)\dots A^n_{\la}(t_n)|0>.
$$
For some models this limit is trivial (equals to zero). However
this does not mean that the stochastic limit of such models is
trivial because we can consider the limit at $ \la\to 0 $ of the
chronologically ordered correlation functions (Green's functions)
$$
 <0|T\left(A^1_{\la}(t_1)\dots A^n_{\la}(t_n)\right)|0>.
$$
Here the $ T $-product is defined as
$$
 T\left(A^1_{\la}(t_1)\dots A^n_{\la}(t_n)\right)=
 A^{i_1}_{\la}(t_{i_1})\dots A^{i_n}_{\la}(t_{i_n})
$$
if $ t_{i_1}\ge\dots\ge t_{i_n} $.

For the non smeared collective operators $a^\pm_\la(k,t)$ from Theorem~\ref{limit1} one has the
following expression for Wigthman functions
$$
 <0|a_\la(p,t)a^+_\la(k,\tau)|0>=
 \frac{1}{\la^2}e^{i(\tau-t)\om(p)/\la^2}\dl(p-k)
$$
$$
  =2\pi\dl(\tau-t)\dl(\om(p))\dl(p-k)+
  \la^22\pi i\dl'(\tau-t)\dl'(\om(p))\dl(p-k)+o(\la^2).
$$
On the analogy with Theorem~\ref{limit1} it can be proved that
$$
 \frac{1}{\la^2}\Theta(t)e^{itx/\la^2}=i\dl(t)\frac{1}{x+i0}-
 \la^2 \dl'(t)\frac{d}{dx}\frac{1}{x+i0}+o(\la^2).
$$
Therefore one has for Green's functions
$$
 <0|T\left(a_{\la}(p,t)a^+_{\la}(k,\tau)\right)|0>=
 \frac{1}{\la^2}\Theta(\tau-t)e^{i(\tau-t)\om(p)/\la^2}\dl(p-k)
$$
$$
 =i\dl(\tau-t)\frac{1}{\om(p)+i0}\dl(p-k)+
 \la^2\dl'(\tau-t)\frac{1}{(\om(p)+i0)^2}\dl(p-k)+o(\la^2).
$$
The first term in this expansion was obtained in~\cite{alv}.


\section{A System Interacting with a Reservoir}\noindent
In the last sections we derive a generalized quantum stochastic
differential equation for corrections to the stochastic limit of
the evolution operator $U(t)$. We will apply the dipole noise
operators described above for study of the higher order
corrections to the stochastic limit. Then we will bring these
equations to the normally ordered form and rewrite it in the form
of generalized quantum stochastic differential equations.

The study of the evolution operator $ U(t) $ describing a system
with interaction is very important in statistical physics and
quantum field theory.

One of the methods of investigation of behavior of quantum system
interacting with a reservoir is the stochastic limit method (see
\cite{alv} for details). One considers quantum models with small
coupling constant $ \la $ for large time $ t $. The idea of the
method is in systematic application of special limiting procedure
$ t\to\infty $ and $ \la\to 0 $ such that $ \la^2t=\tau=Const $.
The result of such limiting procedure is being called the $
\la^2t{\rm -limit} $ or stochastic limit. The limiting equation
for the evolution operator is a quantum stochastic differential
equation.

Using this limiting procedure one can study dynamics of real
physical system with small but finite coupling constant for large
time $ t\sim \tau/\la^2 $ without perturbation theory. For some
important models dynamics in the stochastic limit approximation
became integrable.

The first rigorous result about interaction of a system with a
reservoir where the role of $ \la^2t $-rescalling begun emerge is
due to Bogoliubov~\cite{Bogol1}. Friedrichs, in the context of the
now well known Friedrichs model~\cite{Friedrichs}, was lead to
consider the scaling limit
$$
  t\to\infty,\qquad\la\to 0,\qquad \la^2t=\tau=Const
$$
by second order perturbation theory. This rescaling was used to
derive the master equation by Van Hove~\cite{vh5},
Prigogine~\cite{pr}.

The stochastic limit method leads not only to the master equation
but also to the equation for the dynamics of the reservoir,
see~\cite{alv,akv,apv}.

We will apply the dipole noise operators described above for study
of the higher order corrections to the stochastic limit. We will
consider the quantum model of $ N $-level system (atom)
interacting with a boson field.

Let us describe more concretely this model. Let ${\cal F}$ as
above be the Boson Fock space over the one--particle Hilbert space
$L^2(\mathbb R^d)$. Let ${\mathbb C}^N  $ be the Hilbert space of
$ N $--dimensional complex vectors. Then
$$
\Phi={\mathbb C}^N\otimes {\cal F}
$$
is the Hilbert space of the model which we consider.

The total Hamiltonian of the model $ H_\la $ is (it acts in $ \Phi
$):
\begin{equation}\label{ham}
  H_{\lambda} = H_0+{\lambda}V.
\end{equation}
Here $ H_0={\bf 1}\otimes H_{0R}+H_{0S}\otimes {\bf 1} $ is the
free Hamiltonian with
\begin{equation}\label{H0R}
 H_{0R}=\int\om(k)a^+(k)a(k)dk \, ; \qquad {k\in{\mathbb R}^d}
\end{equation}
($ d=3 $ in the physical space), $ \la $ is a coupling constant,
and the interaction $ V $ is given by
\begin{equation}\label{Vint}
 V=i(D\otimes a^+(g)-D^+\otimes a(g)).
\end{equation}
Here the operators $a^+(g)$ and $a(g)$ describe creation and annihilation of bosons; $ D $ and $
D^+ $ are $ N\times N $- matrices describing transitions between atom levels. The function $g(k)$
is a formfactor (complex valued test function) describing  the interaction of the $ N $-level
system with the reservoir. This Hamiltonian has being applied in quantum optics, tunnelling
processes etc (see~\cite{Cohen,Walls} for details).

The function $ \om(k) $ is a dispersion law. We suppose that the
equation $ \om(k)=\om_0 $ determines regular surface in $ {\mathbb
R}^d $ ($ \om_0\in{\mathbb R} $ will be defined below).  For
example $\om(k)=\sqrt{k^2+m^2}$ for massive bosons ($m$ is the
mass of boson) etc.

Dynamics of the system with $ H_{\la} $ is being determined by the evolution operator $
U_{\la}(t)=e^{itH_0}e^{-itH_{\la}} $. It is the solution of the equation (see~\cite{bogol2}):
\begin{equation}\label{inteq}
  U_{\la}(t)=1-i{\la}\int\limits_0^tV(t')U_{\la}(t')dt'.
\end{equation}
Here $ V(t)=e^{itH_0}Ve^{-itH_0} $ is the interaction operator in
the interaction picture.

We will consider two particular types of the model described
above. For the first type $N$ is arbitrary natural number and the
matrices $ D $ and $ D^+ $ are connected with $ H_{0S} $ by the
following conditions for some $ \om_0\in{\mathbb R} $ (the so
called rotating wave approximation condition):
$$
   e^{itH_{0S}}De^{-itH_{0S}}=De^{-it\om_0}
$$
$$
   e^{itH_{0S}}D^+e^{-itH_{0S}}=D^+e^{it\om_0}.
$$

Under these conditions one immediately has
\begin{equation}\label{V(t)1}
  V(t)=i(D\otimes A^+(t)-D^+\otimes A(t))
\end{equation}
with the time dependent creation and annihilation operators
$$
 A^+(t)=\int g(k)e^{i(\om(k)-\om_0)t}a^+(k)dk,
 \qquad A(t)=\int\bar g(k)e^{-i(\om(k)-\om_0)t}a(k)dk.
$$
After time rescalling $t\to t/\la^2 $ these operators according to
results of the sections~3 and~4 have an asymptotic expansion
\begin{equation}\label{collopexp}
 A(t/\la^2)=\la b(t)+\la^2c(t)+o(\la^2)
\end{equation}
with white and dipole noise satisfying the (causal) commutation
relations which follow from Remark~\ref{remdlp}. These relations
are
\begin{equation}\label{avcr}
 [b(t),b^+(\tau)]=\gamma_0\dl_+(\tau-t)
\end{equation}
\begin{equation}\label{avcrc}
 [c(t),c^+(\tau)]=\gamma_1\dl'_+(\tau-t)
\end{equation}
and other commutators are equal to zero. Here $ \gamma_0 $ and $ \gamma_1 $ are complex numbers
$$
 \gamma_0=-i\int dk\frac{|g(k)|^2}{\om(k)-\om_0-i0}=
 \int\limits_{-\infty}^0d\sigma\int dke^{i\sigma(\om(k)-\om_0)}|g(k)|^2
$$
$$
 \gamma_1=-\int dk\frac{|g(k)|^2}{(\om(k)-\om_0-i0)^2}=
 -\int\limits_{-\infty}^0d\sigma\sigma\int
 dke^{i\sigma(\om(k)-\om_0)}|g(k)|^2.
$$
Usual commutation relations follow from Theorem~\ref{limit1}:
\begin{equation}\label{avcr1}
 [b(t),b^+(\tau)]=\tilde\gamma_0\dl(\tau-t)
\end{equation}
\begin{equation}\label{avcru}
 [c(t),c^+(\tau)]=\tilde\gamma_1\dl'(\tau-t)
\end{equation}
with
$$
 \tilde\gamma_0=\int\limits_{-\infty}^{\infty}d\sigma\int dke^{i\sigma(\om(k)-\om_0)}|g(k)|^2
$$
$$
 \tilde\gamma_1=-\int\limits_{-\infty}^{\infty}d\sigma\sigma\int
 dke^{i\sigma(\om(k)-\om_0)}|g(k)|^2.
$$

The second type of the model is the spin-boson model. This model
corresponds to the case $ N=2$ and no rotating wave approximation
condition is assumed. The spin-boson Hamiltonian in the simple but
non-trivial case $ \e=0 $ has the form (see~\cite{akv,Legett})
\begin{equation}\label{Spb}
  H_\la=-\frac{1}{2}\Delta\sigma_x\otimes 1+1\otimes H_{0R}+
  \la\sigma_z\otimes (a(g)+a^+(g)).
\end{equation}
Here the free Hamiltonian of the boson field $H_{0R}$ is given
by~(\ref{H0R}), $ \sigma_x, \sigma_z $ are the Pauli's matrices,
and $ \D>0 $ is a positive number. This model is widely used in
physics and chemistry and for example describes a dynamical model
of two-level system coupled to an environment
(see~\cite{akv,Legett} for example).

For both these models we study asymptotical behavior of the
evolution operator $ U_\la(t/\la^2) $ with rescalled time. So we
study the expansion
\begin{equation}\label{U(la,t)}
  U(\la,t):=U_\la(t/\la^2)=U_0+\la U_1(t)+o(\la^2).
\end{equation}

It was established~\cite{alv} that $U_0$ satisfies the quantum
stochastic differential equation. White noise operators play an
important role in this consideration. In the present article we
show that higher order correction to $U_0$ which is $U_1$,
satisfies the generalized quantum stochastic differential
equation. In the derivation of this equation an important role
plays the quantum dipole noise.


\section{Higher Order Corrections to the Stochastic Limit of the
Vacuum Expectations}
\noindent Let us write some results for vacuum
expectation of the evolution operator for the models described in
previous section.

In~\cite{AV} Hamiltonians with polynomial self-interaction have
been considered. In our case if the system operators $D$ and $D^+$
are complex numbers then the expression for the vacuum expectation
(ABC-formula) has the form
$$
 <U_{\la}(t/\la^2)>=\exp\bigl(At+\la^2B+\la^2C(t/\la^2)\bigr).
$$
For the linear model, when $D$ and $D^+$ are numbers, one has
$$
 A=-\gamma_0D^+D,\qquad  B=\gamma_1D^+D
$$
$$
 C(t)=D^+D\int
 \frac{|g(k)|^2}{(\om(k)-\om_0-i0)^2}e^{-i(\om(k)-\om_0)t}dk
$$
and $C(t)\to 0$ as $t\to\infty$. Therefore
$\la^2C(t/\la^2)=o(\la^2)$ and
\begin{equation}\label{1stmodelABC}
 <U_{\la}(t/\la^2)>=e^{-\gamma_0tD^+D}(1+\la^2\gamma_1D^+D+o(\la^2)).
\end{equation}

Now let us consider the case then $D$ and $D^+$ are arbitrary, not
necessarily commutative, system operators. Such models were
described in the previous section. The expressions for vacuum
expectations of the evolution operator for such models were
obtained in~\cite{P} by direct calculations. For the model with
rotating wave approximation one has
$$
 <U_{\la}(t/\la^2)>=
 e^{-\gamma_0tD^+D}\left[1+\la^2\gamma_1D^+D
 (1-\gamma_0tD^+D)\right]
$$
\begin{equation}\label{1stmodel}
 -\la^2\gamma_1\sum\limits_{k=1}^{\infty}
 \frac{(-\gamma_0t)^k}{k!}\sum\limits_{p=1}^k(D^+D)^{p-1}D^{+2}D^2(D^+D)^{k-p}+
 o(\la^2).
\end{equation}
In~(\ref{1stmodel}) and~(\ref{Spbvac}) $<\cdot>$ means the
averaging only over the boson vacuum so that the result is an
operator acting in the system Hilbert space $\mathbb C^N$. Note
that if $D$ and $D^+$ are numbers then the series
in~(\ref{1stmodel}) can be summarized and we
obtain~(\ref{1stmodelABC}).

For the spin-boson model one has
$$
 <U(t/\la^2)>_{vac}=e^{iA_1t}\left[1-\la^2(B_1+C_1t)\right]DD^+
$$
\begin{equation}\label{Spbvac}
 +e^{iA_2t}\left[1-\la^2(B_2+C_2t)\right]D^+D+o(\la^2)
\end{equation}
with constants ($l=1,2$)
$$
 A_l=\int dk\frac{|g(k)|^2}{\om_l(k)-i0},\qquad
 B_l=\int dk\frac{|g(k)|^2}{(\om_l(k)-i0)^2},
$$
$$
 C_l=iA_lB_l-i\int dk_1dk_2\frac{|g(k_1)|^2|g(k_2)|^2}
 {(\om_l(k_1)-i0)(\om(k_1)+\om(k_2))}
 \left[\frac{1}{\om_l(k_1)-i0}+ \frac{1}{\om_l(k_2)-i0}\right].
$$
Here $ \om_1(k)=\om(k)-\D,\ \om_2(k)=\om(k)+\D>0 $. The matrices
$D$ and $D^+$ in this case are $2\times 2$-matrices and will be
defined in section~9.


\section{The Normally Ordered Form of
          the Equation for the Evolution Operator}
\noindent Let us write the formal equation for the evolution
operator~(\ref{U(la,t)}) with the interaction
Hamiltonian~(\ref{V(t)1}) using the expansion~(\ref{collopexp}) of
the collective operators. Keeping terms of $O(\la)$ one has
\begin{equation}\label{bno}
 \frac{\partial U(\la,t)}{\partial t}=(D(b^+(t)+\la c^+(t))-D^+(b(t)+\la
 c(t)))U(\la,t)+\dots
\end{equation}
Here operators $b(t)$ and $ c(t) $ are the white and dipole noise
with commutation relations~(\ref{avcr}).

Let us expand the time rescalled evolution operator as the series at the coupling constant
$$
 U(\la,t)=U_\la(t/\la^2)=U_0(t)+\la U_1(t)+\dots
$$
Then comparing terms of the same order of $ \la $ in the LHS and
RHS of equality~(\ref{bno}) one obtains a system of stochastic
differential equations
\begin{equation}\label{U0}
 \frac{d}{dt}U_0(t)=(Db^+(t)-D^+b(t))U_0(t)
\end{equation}
\begin{equation}\label{U1}
 \frac{d}{dt}U_1(t)=(Dc^+(t)-D^+c(t))U_0(t)+(Db^+(t)-D^+b(t))U_1(t).
\end{equation}

The normally ordered form of equations of such kind is the form in
which creation operators are on the left side of the evolution
operator and annihilation operators are on the right one. Such
form of the equation is convenient for calculation of matrix
elements of the evolution operator on coherent (exponential)
vectors. The Eq.~(\ref{U0}) and Eq.~(\ref{U1}) are equivalent to
the following
$$
 \frac{d}{dt}U_0(t)=Db^+(t)U_0(t)-D^+U_0(t)b(t)-D^+[b(t),U_0(t)]
$$
$$
 \frac{d}{dt}U_1(t)=Dc^+(t)U_0(t)-D^+U_0(t)c(t)+Db^+(t)U_1(t)-D^+U_1(t)b(t)
$$
$$
  -D^+[c(t),U_0(t)]-D^+[b(t),U_1(t)].
$$
Therefore in order to bring the Eq.~(\ref{U0}) and Eq.~(\ref{U1})
to the normally ordered form one needs to compute the commutators
$[b(t),U_0(t)]$, $[c(t),U_0(t)]$ and $[b(t),U_1(t)]$.

The first commutator had been computed in~\cite{alv} and equal to
$$
 [b(t),U_0(t)]=\gamma_0DU_0(t)
$$
The second equal to zero because white noise commutes with the
dipole noise and $ U_0(t) $ "consists" only of the white noise.
The last commutator can be computed using causal commutation
relations~(\ref{avcr}), time consecutive principle and integral
form of the equation~(\ref{U1})
$$
 [b(t),U_1(t)]=D\int\limits_0^tdt_1[b(t),b^+(t_1)]U_1(t_1)=\gamma_0DU_1(t).
$$
Using these commutators one immediately obtains
\begin{equation}\label{pdU0}
 \frac{d}{dt}U_0(t)=Db^+(t)U_0(t)-D^+U_0(t)b(t)-\gamma_0D^+DU_0(t)
\end{equation}
$$
 \frac{d}{dt}U_1(t)=Dc^+(t)U_0(t)-D^+U_0(t)c(t)
$$
\begin{equation}\label{pdU1}
 +Db^+(t)U_1(t)-D^+U_1(t)b(t)
 -\gamma_0D^+DU_1(t).
\end{equation}

We can rewrite Eq.~(\ref{pdU0}) and Eq.~(\ref{pdU1}) in the
forms~(\ref{dU0(t)}) and~(\ref{dU1(t)}) correspondingly. The
equation in the form~(\ref{pdU1}) we call the generalized quantum
stochastic differential equation. It is an important task to
develop a theory of equations of such form and then to prove the
existence and uniqueness of the solution. An interesting problem
is to study the unitarity of the solution.


\section{Spin-Boson Hamiltonian}\label{sectionSpb}\noindent
Let us consider the spin-boson model with Hamiltonian~(\ref{Spb}).
The free evolution of $ \sigma_z $ is
$$
 \sigma_z(t)=e^{itH_s}\sigma_ze^{-itH_s}=e^{it\D}D+e^{-it\D}D^+
$$
with
$$
 D=\frac{1}{2}\left(
 \begin{array}{cc}
   1& 1\\
  -1& -1
 \end{array}
 \right)
$$
It is easy to verify that $ D^2=D^{+2}=0$ and $ (DD^+)^2=DD^+ $.
For this interaction (see~\cite{akv})
$$
 V(t)=D\otimes W^+(t)+D^+\otimes W(t).
$$
Here $ W(t)=A_1^+(t)+A_2(t)$ with
$$
 A_l(t)=\int dk\bar g(k)e^{-it\om_l(k)}a(k),\qquad l=1,2
$$
and $ \om_1(k)=\om(k)-\D$, $ \om_2(k)=\om(k)+\D>0 $.

We will use the following expansion for the vacuum expectation
value of\\ $ U(\la,t)\equiv U_{\la}(t/\la^2)$:
$$
 <U(\la,t)>=<U_0(t)>+\la^2<U_2(t)>+o(\la^2).
$$

\begin{theorem}\it
The first correction $ f(t)\equiv <U_2(t)> $ satisfies the equation
\begin{equation}\label{Spbeq}
 \frac{df(t)}{dt}=(iA_1DD^++iA_2D^+D)f(t)-
 DD^+C_1e^{iA_1t}-D^+DC_2e^{iA_2t},
\end{equation}
with constants ($l=1, 2$)
$$
 A_l=\int dk\frac{|g(k)|^2}{\om_l(k)-i0},\quad
 B_l=\int dk\frac{|g(k)|^2}{(\om_l(k)-i0)^2},\quad
 C_l=iA_lB_l-iZ_l,
$$
$$
 Z_l=\int dk_1dk_2\frac{|g(k_1)|^2|g(k_2)|^2}
 {(\om_l(k_1)-i0)(\om(k_1)+\om(k_2))}
 \left[\frac{1}{\om_l(k_1)-i0}+ \frac{1}{\om_l(k_2)-i0}\right]
$$
\end{theorem}

\noindent{\bf Proof.} One has
$$
  \frac{\partial <U(\la,t)>}{\partial t}=-\frac{i}{\la}<V(t/\la^2)U(\la,t)>
$$
$$
  =-i<\left[\left(\frac{1}{\la}A_1(t/\la^2)\otimes D+
  \frac{1}{\la}A_2(t/\la^2)\otimes D^+\right),U(\la,t)\right]>
$$
$$
  =-\frac{1}{\la^2}\int\limits_0^tdt'\biggl\{
  \Bigl[A_1(t/\la^2),A_1^+(t'/\la^2)\Bigr]\otimes DD^+<U(\la,t')>
$$
$$
  +<A_2(t'/\la^2)A_1(t/\la^2)\otimes DD^+U(\la,t')>\biggr\}
$$
$$
  -\frac{1}{\la^2}\int\limits_0^tdt'\biggl\{
  \Bigl[A_2(t/\la^2),A_2^+(t'/\la^2)\Bigr]\otimes D^+D<U(\la,t')>
$$
\begin{equation}\label{Neq}
  +<A_1(t'/\la^2)A_2(t/\la^2)\otimes D^+DU(\la,t')>\biggr\}.
\end{equation}
Let us consider the second term in the first curly bracket. Using
the expression
$$
 U(\la,t)=1-\frac{i}{\la}\int\limits_0^tV(t_1/\la^2)dt_1+
 \left(-\frac{i}{\la}\right)^2\int\limits_0^tdt_1\int\limits_0^{t_1}dt_2
 V(t_1/\la^2)V(t_2/\la^2)U(\la,t_2)
$$
one gets
$$
 \lim\limits_{\la\to 0}-\frac{1}{\la^4}
 \int\limits_0^tdt_1<A_1(t/\la^2)A_2(t/\la^2)\otimes DD^+U(\la,t')>
$$
$$
 =\lim\limits_{\la\to 0}
  \frac{1}{\la^6}\int\limits_0^tdt_1\int\limits_0^{t_1}dt_2
  \int\limits_0^{t_2}dt_3<A_1(t/\la^2)A_2(t_1/\la^2)
  W^+(t_2/\la^2)W(t_3/\la^2)
$$
$$
  \otimes (DD^+)^2U(\la,t_3)>.
$$
We suppose that this limit is equal
$$
 \lim\limits_{\la\to 0}
  \frac{1}{\la^6}\int\limits_0^tdt_1\int\limits_0^{t_1}dt_2
  \int\limits_0^{t_2}dt_3<A_1(t/\la^2)A_2(t_1/\la^2)
  A_2^+(t_2/\la^2)A_1^+(t_3/\la^2)>
$$
$$
  \otimes (DD^+)^2<U_0(t_3)>
$$
$$
 =(DD^+)^2\lim\limits_{\la\to 0}\lim\limits_{\e\to +0}
  \int dk_1\int dk_2|g(k_1)|^2|g(k_2)|^2
  \int\limits_0^{t/\la^2}dt_1
  \int\limits_0^{t_1}dt_2
  \int\limits_0^{t_2}dt_3
  e^{-i\om_1^-(k_1)t/\la^2-i\om_2(k_2)t_1}
$$
$$
 \times\biggl\{e^{i\om_1^-(k_1)t_3+i\om_2(k_2)t_2}+
  e^{i\om_1^-(k_2)t_3)+i\om_2(k_1)t_2}\biggr\}<U_0(t_3)>=
  iDD^+Z_1<U_0(t)>.
$$
Here we denote $\om^-_1(k):=\om_1(k)-i\varepsilon$.

Using Theorem~\ref{limit2} one gets for the first term in the
first curly bracket in~(\ref{Neq})
$$
 -\frac{1}{\la^2}\int\limits_0^t\Bigl[A_1(t/\la^2),A_1^+(t'/\la^2)
 \Bigl]\otimes DD^+<U(\la,t')>dt'
$$
$$
 =iA_1DD^+<U_0(t)>+i\la^2A_1DD^+f(t)-
 \la^2DD^+B_1\frac{d<U_0(t)>}{dt}+o(\la^2).
$$
Making similar computation with the second curly bracket in~(\ref{Neq}) one gets
$$
 \frac{d<U_0(t)>}{dt}+\la^2\frac{df(t)}{dt} =
 (iA_1DD^++iA_2D^+D)<U_0(t)>
$$
$$
 +i\la^2(A_1DD^++A_2D^+D)f(t)
 -\la^2DD^+C_1e^{iA_1t}-\la^2D^+DC_2e^{iA_2t}.
$$

Let us compare the terms before equal degrees of $ \la $ in both
sides of the last relation. For zero order term one has
$$
  \frac{d<U_0(t)>}{dt} = (iA_1DD^++iA_2D^+D)<U_0(t)>\Rightarrow
$$
$$
   <U_0(t)> = e^{iA_1t}DD^++e^{iA_2t}D^+D.
$$
Now we substitute this solution for $<U_0(t)>$ in the terms before
$ \la^2 $ and obtain the equation
$$
 \frac{df(t)}{dt} = (iA_1DD^++iA_2D^+D)f(t)-
 DD^+C_1e^{iA_1t}-D^+DC_2e^{iA_2t}.
$$
This finishes the proof of the theorem.\hspace{6.5cm}\qed

It is easy to verify that a solution of equation~(\ref{Spbeq}) is
$$
 f(t) = [\tilde B_1-C_1t]e^{iA_1t}DD^++[\tilde B_2-C_2t]e^{iA_2t}D^+D.
$$
Here constants $ \tilde B_1$, $\tilde B_2 $ are determined by
initial conditions. In order to obtain~(\ref{Spbvac}) one should
set $ \tilde B_1 = -B_1 $, $ \tilde B_2 = -B_2 $.


\section{Conclusions}\noindent
In the present paper we have considered the operator valued
distribution with commutation relations proportional to the
derivative of $\dl$-function, the so-called dipole noise
operators. We have constructed a representation of these
commutation relations in a Fock space with indefinite metric, i.e.
in a pseudo-Hilbert space. An interesting problem is to construct
a classical analog of this quantum dipole and, more generally,
multipole noise.

The multipole noise is interesting not only as a mathematical
object but it also plays an important role for applications in
physics. In particular, we apply the dipole noise operators to
study some models of quantum optics. For the evolution operator of
these models we obtain stochastic equations which we call
generalized quantum stochastic differential equations. An
important open problem is to develop a theory of these generalized
quantum stochastic equations, including proving of existence and
uniqueness of the solution. An interesting question is to study an
analogue of the unitarity of the solution.

The multipole noise describes higher order corrections to the
stochastic limit of quantum theory. This is an approximate
non-perturbation method which one can apply to study the dynamics
of quantum system interacting with a reservoir in the case when
coupling constant or density of particles of reservoir is a small
parameter and for large time. Therefore it is an important task to
apply this approach to other models of quantum theory.

\nonumsection{Acknowledgment}\noindent

This work is partially supported by the INTAS 99-00545 for I.~V.
and by the INTAS 01/1-200 for A.~P. and also by the RFFI
02-01-01084 and the grant of the leading scientific school
00-15-96073.

\nonumsection{References}

\end{document}